\documentclass[twocolumn]{aastex63}
\usepackage{color}
\usepackage{amsmath}

\newcommand{\ie}{i.e.,\ }
\newcommand{\eg}{e.g.,\ }
\newcommand{\etal}{et~al.\ }
\newcommand{\kms}{km~s$^{-1}$}
\newcommand{\magsec}{mag arcsec$^{-2}$}
\def\brackmueg{$\langle \mu_g \rangle_e$}
\def\rperi{$R_{\rm peri}$}
\def\tperi{$t_{\rm peri}$}

\defcitealias{Mei07}{M07}
\defcitealias{Blakeslee09}{B09}

\begin{document}

\title{The Distance and Dynamical History of the Virgo Cluster Ultradiffuse Galaxy VCC~615}
\shorttitle{Distance and Dynamical History of VCC~615}
\shortauthors{Mihos \etal}

\author{J. Christopher Mihos}
\affiliation{Department of Astronomy, Case Western Reserve University, Cleveland OH 44106, USA}

\author{Patrick R. Durrell}
\affiliation{Department of Physics, Astronomy, Geophysics and Environmental Sciences, Youngstown State University, Youngstown, OH 44555 USA}

\author{Elisa Toloba}
\affiliation{Department of Physics, University of the Pacific, Stockton, CA 95211, USA}

\author{Patrick C\^ot\'e}
\affiliation{National Research Council of Canada, Herzberg Astronomy and Astrophysics Program, Victoria, BC V9E 2E7, Canada}

\author{Laura Ferrarese}
\affiliation{National Research Council of Canada, Herzberg Astronomy and Astrophysics Program, Victoria, BC V9E 2E7, Canada}

\author{Puragra Guhathakurta}
\affiliation{UCO/Lick Observatory, University of California Santa Cruz, Santa Cruz, CA 95064, USA}

\author{Sungsoon Lim}
\affiliation{Department of Astronomy, Yonsei University, 50 Yonsei-ro Seodaemun-gu, Seoul, 03722, Republic of Korea}

\author{Eric W. Peng}
\affiliation{Department of Astronomy, Peking University, Beijing 100871, People's Republic of China}
\affiliation{Kavli Institute of Astronomy and Astrophysics, Peking University, Beijing 100871, People's Republic of China}

\author{Laura V. Sales}
\affiliation{Department of Physics and Astronomy, University of California, Riverside, CA 92521, USA}

\begin{abstract}

We use deep {\sl Hubble Space Telescope} imaging to derive a distance to
the Virgo Cluster ultradiffuse galaxy (UDG) VCC~615 using the tip of the
red giant branch (TRGB) distance estimator. We detect 5,023 stars within
the galaxy, down to a 50\% completeness limit of $\rm F814W \approx
28.0$, using counts in the surrounding field to correct for
contamination due to background sources and Virgo intracluster stars. We
derive an extinction-corrected F814W tip magnitude of $m_{\rm
tip,0}=27.19^{+0.07}_{-0.05}$, yielding a distance of
$d=17.7^{+0.6}_{-0.4}$ Mpc. This places VCC~615 on the far side of the
Virgo Cluster ($d_{\rm Virgo}=16.5$ Mpc), at a Virgocentric distance of 1.3
Mpc and near the virial radius of the main body of Virgo. Coupling this
distance with the galaxy's observed radial velocity, we find that
VCC~615 is on an outbound trajectory, having survived a recent
passage through the inner parts of the cluster. Indeed, our orbit
modeling gives a 50\% chance the galaxy passed inside the Virgo core
($r<620$ kpc) within the past Gyr, although very close passages directly
through the cluster center ($r<200$ kpc) are unlikely. Given VCC~615's
undisturbed morphology, we argue that the galaxy has experienced no
recent and sudden transformation into a UDG due to the cluster
potential, but rather is a long-lived UDG whose relatively wide orbit
and large dynamical mass protect it from stripping and destruction by
Virgo cluster tides. Finally, we also describe the serendipitous
discovery of a nearby Virgo dwarf galaxy projected 90\arcsec\ (7.2 kpc)
away from VCC~615.


\end{abstract}

\keywords{Dwarf Galaxies --- Galaxy clusters --- Galaxy distances --- Galaxy evolution --- Low Surface Brightness Galaxies} 

\section{Introduction}

The wide diversity in properties of different galaxy populations ---
their structure, kinematics, stellar populations, and environments ---
reflects the diversity in their formation and evolutionary histories. Of
particular recent interest are the so-called ``ultradiffuse galaxies''
(UDGs), galaxies with extremely low optical surface brightnesses ($\mu_V
\gtrsim 26$ \magsec) and large effective radii ($r_e > 1.5$ kpc). These
galaxies occupy a range of environments, from galaxy clusters
\citep[\eg][]{vanDokkum15, Mihos15,Koda15} to group and field
environments \citep{Greco18,Barbosa20,Tanoglidis21}. While cluster UDGs
are typically red and lack evidence for star formation
\citep{vanDokkum15}, UDGs in the field span a wide range of colors
\citep{Greco18}, including some which are extremely blue and gas-rich
\citep{Cannon15, Leisman17, Mihos18leo}.

The dark matter content of these galaxies appears similarly diverse.
Dynamical mass estimates of a number of UDGs indicate that they are
extremely dark matter dominated \citep{Beasley16, vanDokkum16, Toloba18,
Forbes21}, and many UDGs possess significant numbers of globular
clusters, more typical of globular cluster systems seen in more massive
hosts \citep{Peng16, vanDokkum17, Lim18, Muller21}. However, direct kinematics
have also revealed that at least some UDGs appear to have a dearth of
dark matter \citep{vanDokkum18, vanDokkum19, Danieli19}, arguing for
fundamental differences between UDGs and the normal galaxy population.

A wide range of formation scenarios have been proposed to explain the
differing properties of UDGs. The simplest possibility is that UDGs
represent the natural extension of normal galaxies down to the lowest
surface brightness, perhaps representing the high-spin tail of the dwarf
galaxy population \citep{Amorisco16}, or low mass galaxies that have
experienced very efficient stellar feedback \citep{diCintio17,Chan18}.
Indeed, there is no discontinuity seen in the distribution of galaxy
surface brightness or size that would mark a well-defined transition
from normal galaxies to UDGs \citep{McGaugh95, Driver05, Lim20}. This,
however, would not explain differences in the dark matter content of
UDGs. An alternative explanation is that UDGs are ``failed'' $L_*$
galaxies \citep{vanDokkum15}, where gas was lost from the system before
a luminous galaxy could form inside an otherwise normal dark halo. These
scenarios need not be exclusive; observational data suggests a large
diversity in kinematics and globular cluster populations for these
diffuse galaxies, better explained by a combination of failed massive
objects and lower mass dwarfs-sized galaxies
\citep[\eg][]{Lim18,Lim20,Doppel21}.

The cluster environment offers additional evolutionary pathways to explain
cluster UDGs. One possibility for cluster UDGs is that they started as
otherwise normal dwarf galaxies, but have been dynamically heated and
``puffed up'' by interactions within the cluster \citep{Moore96, Liao19,
Carleton19, Tremmel20}, or after gas is ram-pressure stripped by the 
intracluster medium \citep{SafScan17}. UDGs that are satellites in
groups and clusters have also been shown to form by tidal stripping
of otherwise normal galaxies either with \citep{Carleton19} or without \citep{Sales20} 
cored dark matter halos. However, not all cluster UDGs may have formed
in response  to the cluster environment; simulations show that a significant
fraction of UDGs found in clusters may have been object ``born'' as UDGs
in the field environment and later accreted into the cluster \citep{Sales20}.

In addition to providing a formation channel for some UDGs,
the cluster environment also raises questions about their
dynamical evolution and survival.   Large, low
density galaxies should be the ones most easily destroyed by close
interactions and the cluster tidal field, yet UDGs appear common in rich
clusters. In the Coma Cluster, UDGs tend to avoid the cluster center
\citep{vanDokkum15}, perhaps reflecting this fragility --- these objects
may be field UDGs falling into the cluster for the first time, or moving
on orbits which avoid the cluster core entirely. However, some UDGs do
exist in the core of both the Coma and Virgo Clusters \citep{Koda15,
Mihos15, Mihos18, Lim20}; in some cases, these show evidence for tidal
stripping and may be examples of UDGs spawned by dynamical heating
within the cluster. These objects may thus be short lived as they are
continually stripped and eventually destroyed by cluster tides. Here
though the question of their dark matter content resurfaces: if UDGs are
cocooned in massive dark halos, they may be more resilient against tidal
destruction and instead be long-lived members of the cluster galaxy
population.

Thus, the story of cluster UDGs is complicated both by the questions of
their origins and of their subsequent evolution within the cluster
environment, and these galaxies are likely a heterogeneous population of
galaxies following a variety of evolutionary paths \citep{Lim18,
Sales20}. To unravel these differences requires a better understanding
of their local environments and their dynamical history. Are the UDGs
found in cluster cores really in the deepest part of the cluster
potential, or are they objects in the outskirts merely projected near
the core? Are cluster UDGs objects falling into the cluster for the
first time, or have they already experienced core passage? Are they on
radial orbits, or orbits which avoid the cluster core and keep the UDGs
confined to the cluster outskirts?

Here we address these questions for the Virgo Cluster UDG VCC~615 by
using deep {\sl Hubble Space Telescope} imaging to obtain an accurate
tip of the red giant branch (TRGB) distance to the galaxy and explore
its position within the cluster. The Virgo Cluster is ideal for a study
such as this; at a distance of only 16.5 Mpc \citep[][hereafter M07 and
B09, respectively]{Mei07,Blakeslee09}, it is close enough and large
enough that distance estimates offering $<$10\% uncertainties can
pinpoint the three-dimensional locations of a galaxy {\it within} the
cluster potential. VCC~615 is included in the \citet{Lim20} catalog of
Virgo UDGs derived from {\sl Next Generation Virgo Cluster Survey}
\citep[NGVS;][]{Ferrarese12} imaging, and has an effective surface
brightness of \brackmueg=26.3 \magsec, effective radius
$r_e=26.3$\arcsec\ (2.1 kpc at $d_{\rm Virgo}$), and total magnitude
$m_g=17.3$ ($M_g=-13.8$). The galaxy lies projected 1.9\degr\ (550 kpc)
SSW of M87, just inside the Virgo Cluster core\footnote{Here we define
the core radius to be the scale radius of the NFW potential describing
the main body of Virgo ($R_s=620$ kpc), as derived by
\citet{McLaughlin99}.}. After deriving the true three-dimensional
position of VCC~615 within Virgo, we couple this information with its
measured line-of-sight velocity ($2094\pm3$ \kms; \citealt{Toloba18}) to
constrain the galaxy's orbit within the cluster, probe its dynamical
history, and compare to scenarios for cluster UDG formation and
evolution.

\section{Observational Data}

\begin{figure}[]
\centerline{\includegraphics[width=3.3truein]{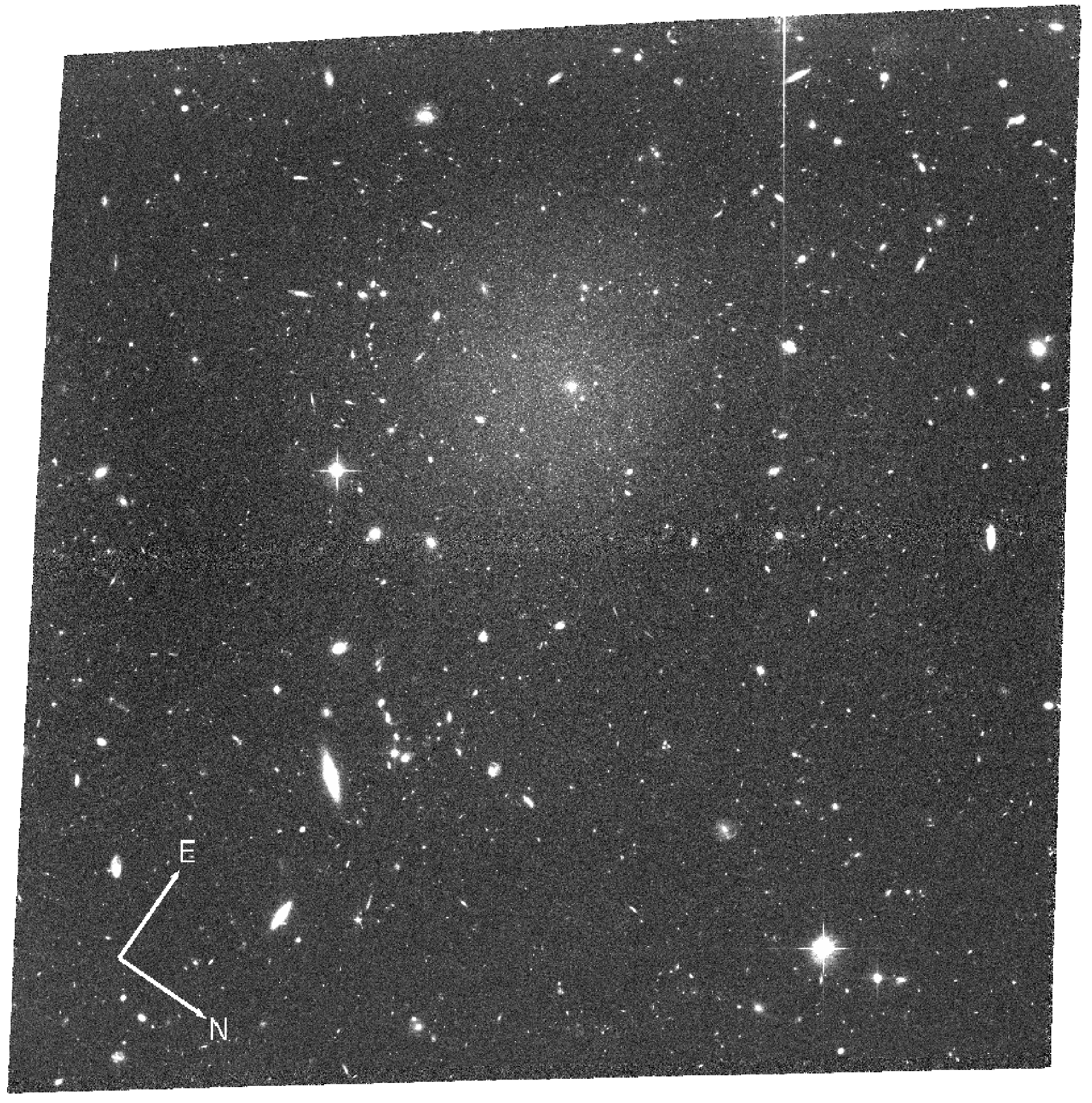}}
\caption{Stacked F814W ACS image of VCC~615, with a total exposure time of 11$\times$1200s.
The field of view is 202\arcsec$\times$202\arcsec, and the Virgo Cluster center (defined
by the giant elliptical M87) is 1.95\degr\ away  (560 kpc in projection) to the east (upper right) of the 
field.}
\label{image}
\end{figure}

We imaged VCC~615 using the Wide Field Channel (WFC) of the Advanced
Camera for Surveys (ACS) on the {\sl Hubble Space Telescope} (HST) as
part of program GO-15258. Figure~\ref{image} shows the imaging field,
centered at $(\alpha,\delta)_{\rm J2000.0}$ = (12:23:02.7,+12:01:10.0)
and offset 30\arcsec\ to the NW of VCC~615 to avoid nearby bright stars
and also to provide sufficient coverage of the surrounding environment
to enable proper background source estimation. The field was imaged with
the F814W filter over 7 orbits in 4 visits, where each orbit consisted
of a pair of 1200s exposures. Each visit made use of a small ($\sim$
3.5--4.5 pixel) custom 4 point box dither pattern to aid in sub-pixel
sampling of the ACS images and the parallel {\sl WFC3} images (to be
discussed in a later paper) and to avoid placing any objects on bad or
hot pixels. Furthermore, the different visits were further shifted in 20
or 60 pixel offsets to cover the ACS chip gap to allow photometry of all
objects within VCC~615.

Because of the lack of sufficient {\sl HST} guide stars near VCC~615, we
were forced to use single star guiding mode for all exposures in our
program. This proved somewhat problematic, as visual inspection of the
14 individual F814W {\it .flc} (CTE-corrected) images revealed that many
showed slightly elliptical PSFs, indicative of a variable level of
trailing during the exposures. From this inspection (and stacking image
subsets using astrodrizzle) we found that 11 of 14 images were of
sufficiently good quality to use for point-source photometry; the
remaining 3 images were discarded. Fortunately (as noted below) the
photometric depth from the remaining 11 images were sufficient to
clearly detect RGB stars in VCC~615 to a depth where we could reliably
get a TRGB-based distance.

Furthermore, the variable trailing also meant that archived F814W
drizzled stacks from each of the 4 visits were also unusable. DOLPHOT (see
next section) requires the use of a sufficiently deep drizzled image as
an astrometric reference, so we created a single deep 11 image drizzled
stack by (a) re-registering all images within each visit using source
positions from Source-Extractor \citep{source-ext} and using the {\tt
tweakreg} and {\tt tweakback} packages within {\tt drizzlepac} to
re-write the WCS for each image, and then (b) re-registering the images
from {\tt all} visits using {\tt tweakreg} to match the WCS of the very
first F814W image from the entire sequence. The package {\tt
astrodrizzle} within {\tt drizzlepac} was then used to create a single,
deep F814W image of the ACS field (shown in Figure~\ref{image}) using 11
exposures with a total exposure time of 13200s.

\subsection{Point Source Photometry}

Photometry on the ACS images was performed with the ACS module of the
DOLPHOT software package \citep[an updated version of
HSTPhot;][]{dol00}, which is designed for point-source photometry of
objects on the individual CTE-corrected {\it flc} images using
pre-computed Tiny Tim PSFs \citep{tinytim}. Object detection and
photometry was performed on all 11 F814W images simultaneously, using
the deep image stack created above as the astrometric standard
(reference) image. We used the most recent version\footnote{Nov. 2019
version; http://americano.dolphinsim.com/dolphot/} of DOLPHOT~2.0 to
pre-process the raw {\it flc} images to apply bad-pixel masks and
pixel-area masks ({\tt acsmask}), split the images into the individual
WFC1/2 chip images ({\tt splitgroups}), and to construct an initial
background sky map for each chip from each image ({\tt calcsky}).

Photometry with DOLPHOT is very dependent on the input parameters used
\citep[see the excellent discussion in][]{Williams14}, so we
experimented with a number of input parameters, finally settling on
values similar to that used in previous deep crowded-field photometric
studies \citep{Williams14,Shen21} and/or suggested values from the
DOLPHOT User's Manual useful for faint, relatively crowded regions of
stellar objects such as that exhibited by VCC~615. More specifically, we
used RAPER=3.0 pixels, the FITSKY=2 option for determining the sky
values from smaller annuli just outside the aperture radius (but within
the adopted PSF size RPSF=13 pixels), and adopting FORCE1=1 so that
DOLPHOT 'forces' objects to be fit as stars. The only changes we made to
the usual DOLPHOT workflow was the derivation of the aperture
corrections on each chip/image. With so few bright stellar objects in
our frames, we found the DOLPHOT computed aperture corrections could be
dominated by a few brighter non-stellar objects, which resulted in a
large spread of $\sim 0.05$ mag in the F814W magnitudes of bright stars
on the individual {\it flc} images. To improve this, we input our own
visually-selected list of 59 stellar objects from which to derive
aperture corrections; while this avoided the above issue, it also meant
that some frames had very few objects from which to determine the
aperture correction, potentially biasing the magnitudes derived from
those images only. Indeed, we found that aperture corrections on images
with $N<8$ aperture stars used differed by those with more aperture
stars by 0.02 mag. Thus the final adopted aperture corrections for each
image/chip are (a) DOLPHOT-computed values for those frames with $N>8$
aperture stars, and (b) a fixed value of $-0.038$ (based on the average
value from DOLPHOT values from the $N>8$ group) for those frames where
$N<8$. This reduced the standard deviation of the individual magnitudes
to less than 0.03 mag. We consider this a lower limit on the total
absolute error in our final, DOLPHOT-combined F814W magnitudes. Finally,
the instrumental magnitudes were converted to the VEGAmag {\sl HST}
photometric system by using a zeropoint of 25.523 for F814W used in the
most recent version of DOLPHOT. We report all photometry in Vega
magnitudes unless explicitly stated otherwise.

\subsubsection{Artificial Stars}

\begin{figure*}[]
\centerline{\includegraphics[width=7.0truein]{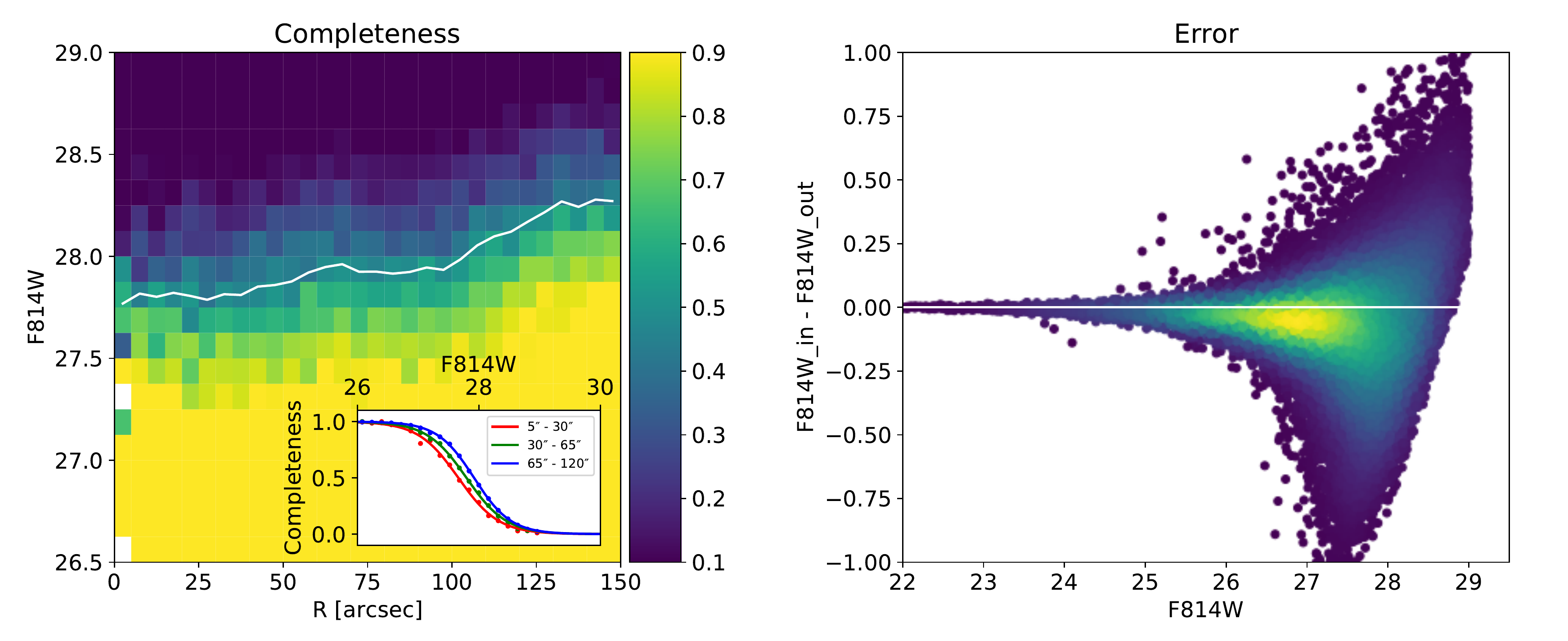}}
\caption{Photometric completeness and uncertainties measured using
artificial star tests. The colormap in the left panel shows completeness as a function
of magnitude and distance from the center of VCC~615, with the white
line showing the 50\% completeness limit as a function of radial
distance. The inset plot in that panel shows the completeness curves extracted
in three different radial ranges.} The right panel shows the absolute error in derived magnitude
(i.e., the difference between input and output magnitudes) as a function
of magnitude.
\label{artstars}
\end{figure*}

We also used DOLPHOT to determine photometric completeness and
uncertainty through artificial star tests. We injected 100,000
artificial stars into the ACS image over a magnitude range $22<{\rm
F814W}< 29$, and use the same cuts on photometric signal-to-noise and
image shape (see below) that we use in extracting the actual photometric
data. Averaged across the entire ACS field, we find a 50\% completeness
limit of F814W=27.9, but as seen in Figure~\ref{artstars}, the exact
value depends somewhat on the distance from the center of VCC~615. 
Because of the galaxy's very low surface brightness, the common
factors contributing to incompleteness --- such as crowding and 
heightened integrated light --- are reduced, such that the radial gradient 
in incompeteness is rather mild. In empty regions far
from VCC~615 (at $R>100$\arcsec) 50\% completeness is found at
F814W=28.1, while in the inner parts of the galaxy (at $R<30$\arcsec)
the limit has risen by less than half a magnitude, to F814W=27.7.
At the average 50\% limit of F814W=27.9 the median photometric error as
reported by DOLPHOT is $\approx$ 0.18 mag, but Figure~\ref{artstars} shows that
the true error (defined as the difference between the injected and
measured magnitudes of the artificial stars) shows a complicated
behavior: at faint magnitudes the scatter in photometric error becomes
asymmetric and biased towards measuring magnitudes slightly fainter than
true (by $\sim$ 0.15 mag at F814W=27.9; see also \citet{Danieli20,
Shen21}). We fold these biases in photometric skew and scatter as a
function of magnitude into our RGB distance estimation in Section~4
below.

\subsubsection{Point Source Selection}

To assemble the final, cleanest photometric sample of stellar sources on
our imaging data, we start by applying a number of photometric and
spatial cuts to the DOLPHOT photometry to avoid spurious objects from
contaminating the sample; this is particularly important since our deep
imaging is only in one filter and thus has no color information to aid
in the selection of RGB stars. We begin by removing sources found within
60 pixels of the edge of the stacked F814W image, as well as any source
detected near the bright column bleed visible in the northeast (right)
side of the images. Next, we mask any source found on or near saturated
stars and bright background galaxies on the image --- these are
typically sources which are either saturated star artifacts or objects
intrinsic to the background galaxy, rather than being RGB stars in
VCC~615 (or the surrounding Virgo environment). We also only select
objects with DOLPHOT object TYPE = 1 (``good star'') and signal-to-noise
$>3.0$, which removes obviously extended sources and sources at low
$S/N$. To remove sources with suspect photometry due to crowding by
nearby objects, we also reject sources with the DOLPHOT CROWD parameter
$>0.2$. Finally, we apply a more rigorous cut on the DOLPHOT sharpness
parameter to select only those objects most likely to be point sources.
Based on visual inspection of objects over a wide range of brightness,
we use a selection function of $|{\rm SHARP}| < 0.1 + 0.6e^{\rm
F814W-28.7}$, similar in form to our previous studies of stellar
populations in the nearby galaxies M81 \citep{Durrell10} and M51
\citep{Mihos18}. When applied to our artificial star tests, this
sharpness cut removes only 2.5\% of the sources brighter than our 50\%
completeness limit, demonstrating that this cut is not overly aggressive
at removing true point sources. After all cuts, the final photometric
sample contains 16,045 sources over the entire ACS field, with 5,023
sources brighter than our 50\% completeness limit of F814W=27.9.

\section{Spatial Distribution of Point Sources}

\begin{figure*}[]
\centerline{\includegraphics[width=7.0truein]{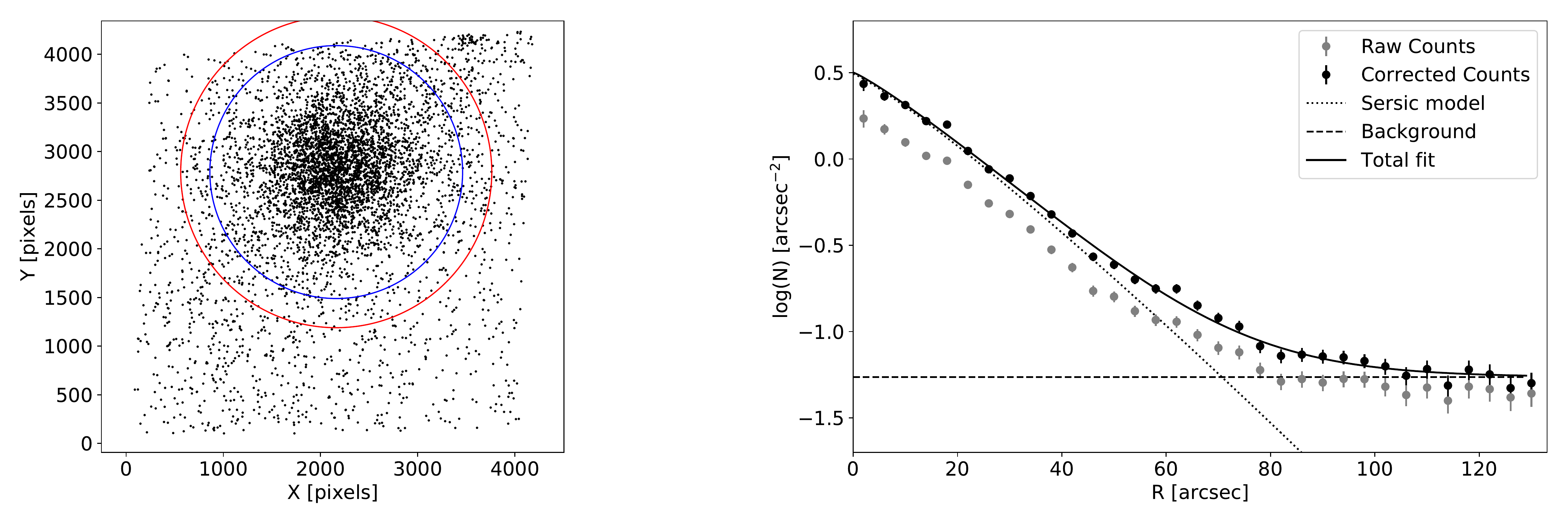}}
\caption{Left: The spatial distribution of detected point sources in our
photometric sample over the magnitude range $26.5 \leq {\rm F814W} \leq
28.0$. The blue circle shows the 65\arcsec radius inside which we draw
our photometric sample of VCC~615 stars, while red circle shows the
80\arcsec radius outside which we measure background contamination.
Right: The radial number density of stellar sources. Grey points show
the observed number density, while black points show the profile
corrected for radially varying incompleteness. The fitted S\'ersic
profile is shown as a black line.}
\label{spatial}
\end{figure*}

Figure~\ref{spatial} shows the spatial distribution of the sources in
our final photometric catalog over the magnitude range $26.5 \leq {\rm
F814W} \leq 28.0$. Overall, the radial decline in density around VCC~615
is clear. On small scales, some regions appear devoid of sources; this
is due to the presence of bright stars or galaxies in that region,
around which we have rejected sources from our catalog. We also note an
overdensity of sources near the upper right corner of the image. These
sources are embedded within a very diffuse envelope of starlight roughly
4\arcsec\ in radius (320 pc at Virgo); this object appears to be a
small, uncataloged, and extremely low surface brightness dwarf galaxy in
which we have resolved its brightest stars. To avoid sources in this
object biasing our estimate of the background source density, we 
mask the object when characterizing populations in the field.

Figure~\ref{spatial} also shows the radial density profile of point
sources around VCC~615, selected in the same magnitude range shown in
Figure~\ref{spatial}a. In this plot, grey points show the raw counts,
while black points show the corrected profile after applying the
radially-dependent incompleteness correction shown in
Figure~\ref{artstars}. The density profile follows a roughly exponential
decline with radius (equivalent to a S\'ersic profile with index $n=1$),
typical for low luminosity spheroidals in the Virgo Cluster. The density
distribution of sources at large distances from VCC~615 show no evidence
of clustering in ways that suggest the presence of low density stellar
streams or tidal distortions of the galaxy. We confirmed this by
smoothing the density field on a variety of spatial scales and looking
at the resulting isodensity contours; in no case was any significant
structure found beyond a radius of $\sim 70$\arcsec. The isodensity
contours of the galaxy appear quite round, consistent with the $g$-band
isophotal analysis of \citet{Lim20}, who found an axial ratio of
($(b/a)=1$) and no obvious tidal features around the galaxy. Fitting the
density profile shown in Figure~\ref{spatial} to a S\'ersic model with
constant background yields an effective radius of $r_e=28.3\arcsec \pm
0.5\arcsec$ and S\'ersic index $n=0.88\pm0.04$. Compared to the $g$-band
integrated light profile from {\sl NGVS} data ($r_e=26.33\arcsec$,
$n=0.67$, \brackmueg=26.86; \citealt{Lim20}), our stellar density
profile is similar in effective radius but shows a somewhat higher
S\'ersic index.

The surface density of contaminating sources in the field surrounding
VCC~615 is likely due to a combination of unresolved background objects
and possible intracluster red giants in the Virgo environment.
Over the magnitude range $27\leq F814W \leq 28.5$, the
logarithmic slope of the counts in the field is $\approx +0.75\ {\rm
dex\ mag^{-1}}$, comparable to that found by \citet{Mihos18} ($+0.6\
{\rm dex\ mag^{-1}}$) for imaging in the parallel fields of the Abell
2744 Hubble Frontier Field data \citep{Lotz17}. However the density of
sources here is roughly three times higher than that found in the
Frontier Field data. Around VCC~615 we find a source density of 0.055
arcsec$^{-2}$ over that same magnitude range, compared to 0.0175
arcsec$^{-2}$ in the Frontier Field. This difference signals the likely
presence of Virgo intracluster stars around VCC~615. While deep imaging
of Virgo detects no intracluster light (ICL) in the field down to a
limiting surface brightness of $\mu_B=29.5$ mag arcsec$^{-2}$
\citep{Mihos17}, star counts are able to trace the ICL even deeper in
equivalent surface brightness. We can use the background source level
around VCC~615 to place an upper limit on the underlying ICL surface
brightness. If we assume all these sources are old (10 Gyr) metal-poor
([M/H]$=-2$) red giants at the Virgo distance, the PARSEC stellar
population synthesis models \citep{Bressan12} yield an equivalent
surface brightness of $\mu_B > 30.1$ mag arcsec$^{-2}$, consistent with
the lack of diffuse broadband light around VCC~615 \citep{Mihos17}.
However, the true ICL surface brightness is likely even lower, since
many of these sources will be true background objects rather than
intracluster stars. If true background objects are present at a level
comparable to that traced by the Frontier Field data, subtracting that
level from the observed number density yields our best estimate for
local surface brightness of the Virgo ICL around VCC~615: $\mu_B = 30.5$
mag arcsec$^{-2}$.

\begin{figure}[]
\centerline{\includegraphics[width=3.2truein]{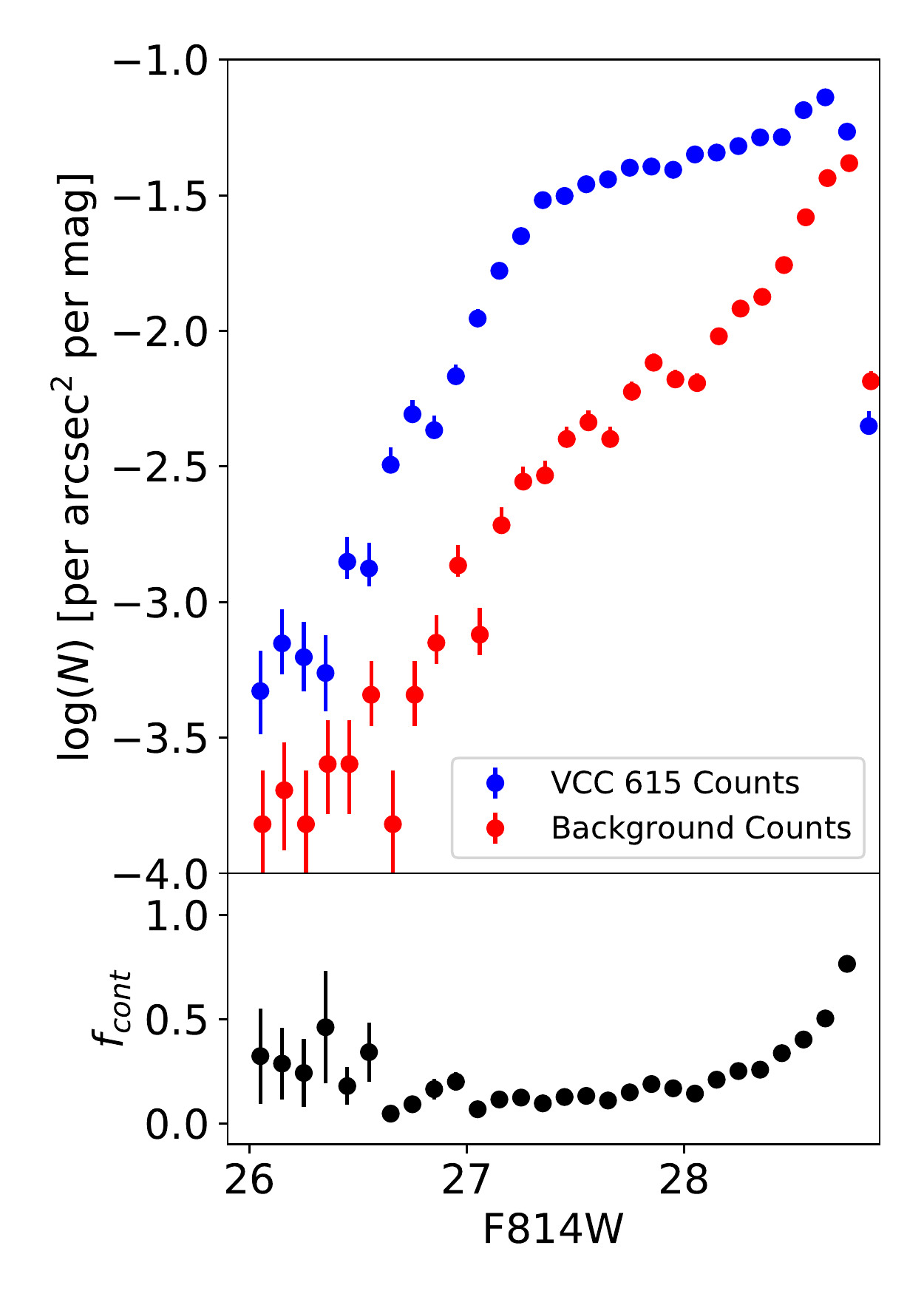}}
\caption{Top: The observed, binned luminosity function of stellar
sources in VCC~615 and the surrounding field (uncorrected for
incompleteness). Bottom: The estimated contamination fraction of
background sources in the VCC~615 luminosity function, defined as
$N_{back}/N_{VCC615}$.}
\label{lumfuncs}
\end{figure}

The spatial distribution of sources in the field also lets us determine
the optimal radial range for our TRGB analysis. Choosing a smaller
radius will reduce background contamination but yield fewer stars for
the photometric sample, while a larger radius will contain more stars
but also make contamination worse. Given the surface density profile
fits shown in Figure~\ref{spatial}b, we choose an outer radius for the
galaxy of $65\arcsec$; enlarging the radius beyond this value adds
background contaminants at a faster rate than VCC~615 stars. The
luminosity function of sources within VCC~615 (i.e., at
$r\leq65\arcsec$) is shown in Figure~\ref{lumfuncs}, as is the
luminosity function of the background sources measured at $r>80\arcsec$,
where the density profile shown in Figure~\ref{spatial} levels off to a
constant background density. The contamination fraction, defined as the
ratio of the surface density of counts in the background to that within
VCC~615, is shown in the bottom of the figure, and is $\approx$ 0.1--0.2
over the magnitude range $26.5 \leq {\rm F814W} \leq 28.0$, climbing to
values $\gtrsim 0.5$ at magnitudes fainter than {\rm F814W}=28.5.

In the VCC~615 luminosity function, the counts show a relatively shallow
slope at fainter magnitudes, but drop off much more quickly at
magnitudes brighter than F814W $\approx$ 27.3, likely marking the
approximate RGB tip. However, even brighter than the tip we find an
excess of stellar sources in VCC~615, suggesting the presence of
asymptotic giant branch (AGB) stars indicative of an intermediate age
population.

\subsection{The nearby LSB dwarf galaxy}

\begin{figure*}[]
\centerline{\includegraphics[width=7.0truein]{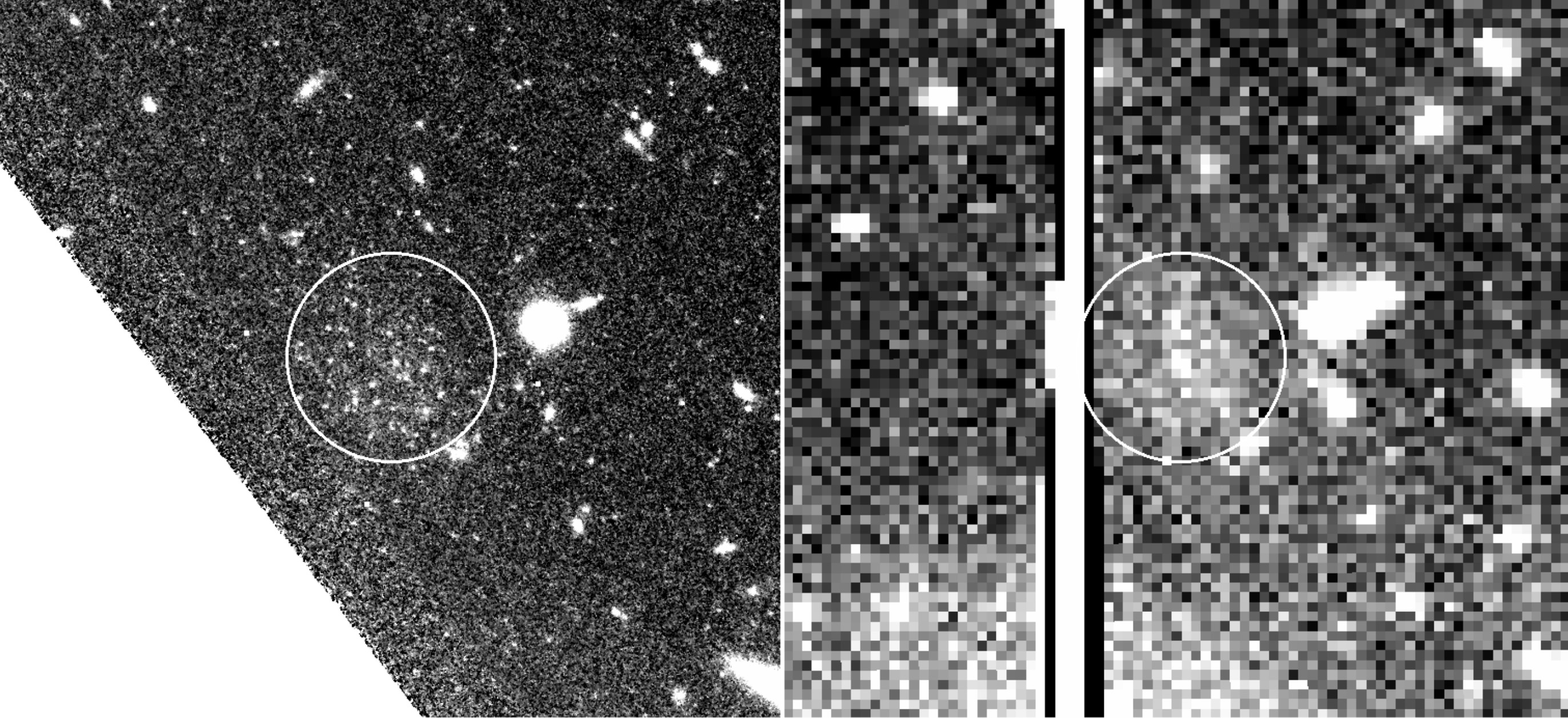}}
\caption{Virgo dwarf galaxy candidate imaged in our ACS data (left) and
in deep NGVS $g$-band imaging (right). North is up, East is to the left,
and the circle has a radius of 4\arcsec. The NGVS image has been
rebinned 2x2 pixels to increase signal to noise, and the vertical column
bleed and scattered light at the bottom of that image are both due to a
bright ($g=12.3$) star 30\arcsec\ to the South.
}
\label{dwarfims}
\end{figure*}

\begin{figure}[]
\centerline{\includegraphics[width=3.2truein]{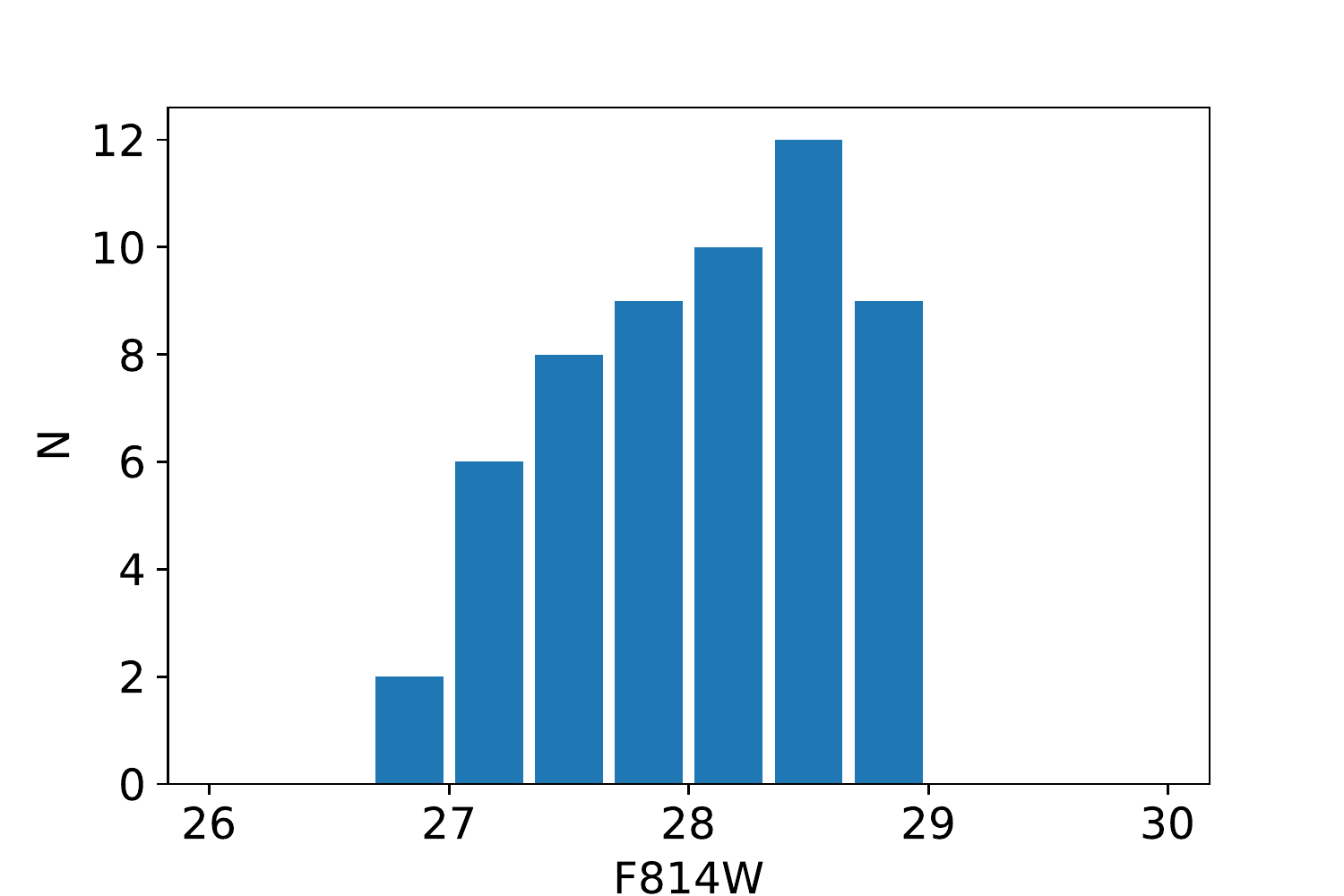}}
\caption{The luminosity function of point sources detected within 
4\arcsec\ of the dwarf galaxy candidate shown in Figure~\ref{dwarfims}, 
in bins of width $\Delta m = 0.33$ mags.}
\label{dwarfLF}
\end{figure}

Before turning to the TRGB analysis of VCC~615, we return briefly to the
dwarf galaxy detected near the upper right (east) edge of our ACS field,
at a position of $(\alpha,\delta)_{J2000.0} =$ (12:23:11.0,
+12:01:10.3). Figure~\ref{dwarfims} shows the ACS imaging of this object
in more detail, along with deep $g$-band imaging of the field from NGVS,
where the object is also clearly visible. The luminosity function of
point sources detected within the 4\arcsec\ is shown in
Figure~\ref{dwarfLF}; while the total number of sources detected (55) is
too small to do robust modeling of the luminosity function, the shape of
the distribution is consistent with what one would expect for RGB stars
at the distance of Virgo: a jump in counts near the expected RGB tip at
F814W $\approx$ 27, with a steep rise towards fainter magnitudes. If we
assume the object is at the Virgo distance of 16.5 Mpc, then after
correction for contamination and incompleteness, inside the $r=4\arcsec$
circle shown in Figure~\ref{dwarfims} we find $N=12^{+2}_{-1}$ sources
between $m_{tip}$ and $m_{tip}+0.5$. Assuming an old (10 Gyr) metal-poor
([M/H]=$-1.5$) stellar population as described by the PARSEC models of
\citet{Bressan12}, these counts would correspond to a total stellar mass
(inside of the 4\arcsec\ radius) of $6.8^{+1.2}_{-0.7} \times 10^5\ {\rm
M_\sun}$ and a $g$-band magnitude of $m_g=23.0^{+0.1}_{-0.2}$.

While the object is clearly detected in the deep NGVS $g$-band imaging,
its small size, low surface brightness, and proximity to a nearby bright
($g=12.3$) star makes detailed surface photometry difficult. Rather than
do a full S\'ersic model of the light profile, we instead fit a simple
exponential model to the profile (\ie S\'ersic $n=1$, typical of low
luminsity spheroids in Virgo), yielding $r_e=5.1\pm0.3\arcsec$, $\langle
\mu_g \rangle_e=27.4\pm0.1$ mag arcsec$^{-2}$, and
$m_{g,tot}=21.8\pm0.1$. This total magnitude is nearly identical to the
50\% completeness limit of the NGVS Virgo galaxy catalog
\citep[$m_{g,50}=22.0$][]{Ferrarese20}; that and the contamination from
the nearby star explain its absence in the NGVS catalog. Inside
$r=4\arcsec$, the object has an integrated magnitude of $m_g=22.9$,
comparable to that derived above using counts near the RGB tip. At the
Virgo distance, these numbers translate to an absolute $g$ magnitude of
$M_g=-9.2$ and a physical effective radius of $r_e=400$ pc, roughly
similar to Milky Way dwarf spheroidals such as Sextans or Ursa Minor
\citep[\eg][]{McConnachie12}. Whether or not this object is physically
associated with VCC~615 is unclear; while it is projected only
90\arcsec\ (7.2 kpc) away, we see no evidence for interaction between
the two, and it may just be a chance projection in the crowded Virgo
Cluster environment.

\section{TRGB Distance Estimation}

In principle, the sharpness of the luminosity function discontinuity at the RGB tip
means that ``edge detectors'' such as the classical Sobel filter can detect it
in a straightforward fashion \citep[see, \eg][]{Sakai96, Madore09, Jang17b,
Beaton18}. In our case, however, a number of
factors complicate such an approach. We have potentially significant
contamination in our sample, due both to the likely presence of intracluster
red giant stars and to the fact that, working in only one filter, we have no color
information to help remove background contaminants.  Furthermore, we are working 
near the limiting depth of the data, where the completeness corrections and asymmetric
photometric errors begin to play important roles in the error distributions.
The combination of these effects means that any discrete break in the underlying
luminosity function ends up being both diluted and smeared asymmetrically, and
becomes very hard to find using classical edge-detector algorithms.

To derive the red giant branch tip magnitude in VCC~615, we instead use a
Bayesian approach similar to that used in previous TRGB studies of old
stellar populations in nearby galaxies (\eg
\citealt{Makarov06,Tollerud16}). In this approach, we model the
underlying stellar luminosity function, convolved with the photometric
errors and incompleteness derived from our artificial star tests, and
with background contamination added based on the LF of sources in the
surrounding environment. This model involves five free parameters
(described below) along with strongly non-Gaussian photometric errors,
making a Bayesian approach ideal for parameter extraction. This
approach also factors in the photometric data all along the luminosity function,
 not just at the location of the break, making it a more robust measure of behavior of the luminosity
function as it crosses the RGB tip.

For the underlying stellar luminosity function (before errors or
contamination), we adopt a broken power law form:

\begin{equation}
\psi(m)=
\begin{cases}
10^{a(m-m_{\rm tip})+b}, & m>m_{\rm tip}\\
10^{c(m-m_{\rm tip})},& m<m_{\rm tip}
\end{cases}
\end{equation}

\noindent where $m_{\rm tip}$ is the magnitude of the RGB tip, $a$ and $c$ are
the power law slopes of the luminosity function fainter and brighter than the RGB tip, respectively, 
and $b$ is the discontinuity in counts at the tip. Choices for these
parameters are described in more detail below.

For a given input luminosity function, we then factor in photometric
errors, incompleteness, and contamination to create an observed
luminosity function:

$$\phi(m)=\int \psi(m^\prime) C(m^\prime) E(m^\prime) dm^\prime + B(m),$$

\noindent where $C$ and $E$ are the photometric incompleteness and
uncertainty models, respectively, and $B(m)$ is the background
contamination model, all described below.

We model the incompleteness using the logistic function:

$$C(m)={1 \over {1+e^{(m-m_{50})/w}}}$$

\noindent where $m_{50}$ is the 50\% completeness limit and $w$
describes how quickly the completeness drops around $m_{50}$. For
simplicity, we do not incorporate a spatially varying incompleteness
model, but rather use a single function where the parameters of the
logistic function are derived from artificial stars located over the
same radial range from which we draw the observed VCC~615 luminosity
function ($r=5-65\arcsec$), yielding $(m_{50},w)=(27.77, 0.33)$.

The photometric error model is more complicated. An examination of
Figure~\ref{artstars}b shows that there is both a systematic shift and
asymmetric spread of the photometric error as a function of magnitude,
and is not well-described by a simple Gaussian model. Instead, we use a
skew-normal distribution \citep{Azzalini09} to model the error function:
$$E(m)=\left({2\over w}\right)G\left({{m-m_0} \over
\omega}\right)F\left(\alpha{{m-m_0} \over \omega}\right),$$

\noindent where $G$ is a normal Gaussian function and $F$ is its
cumulative distribution function. The parameters $m_0, \omega,$ and
$\alpha$ describe the offset, spread, and asymmetry of the skew-normal
distribution, respectively. We bin the artificial stars by magnitude,
using bins of 0.2 mag width, then fit $m_0, \omega,$ and $\alpha$ as a
function of magnitude to capture the change in shift and asymmetry as a
function of magnitude.

Finally, we model background contamination as $B(m)=f_c \times
\rho_b(m)$ where $\rho_b$ is the luminosity function of the background
area shown in Figure~\ref{lumfuncs} and $f_c$ is the ratio of the
background counts to total counts measured at F814W=28.0. Based on the
luminosity functions shown in Figure~\ref{lumfuncs}, $f_c=0.18$, but in
our Bayesian analysis we allow it to vary somewhat to capture the
uncertainty in the overall normalization of the background luminosity
function. We note that unlike the parameterized stellar luminosity
function $\psi$, the background luminosity function comes directly from
the imaging data, so we need not apply the error and incompleteness
models to it --- these effects are implicitly contained in the derived
background luminosity function.

Given a model luminosity function, the log likelihood function is then
given by \citep{Makarov06}:

$$\mathcal{L}=-\sum_{i=1}^N \ln \phi(m_i \vert \mathbf{x}) + N \ln \int_{m_{\rm min}}^{m_{\rm max}} \phi(m\vert \mathbf{x}) dm$$

where {\bf x} represents the five parameters of the model ($m_{\rm tip}, a,
b, c,$ and $f_c$) and the summation is over every star detected in the
magnitude range $(m_{\rm min},m_{\rm max})$. We restrict our analysis to the
magnitude range F814W=(26,28); at brighter magnitudes there are very few
sources, while at fainter magnitudes both incompleteness and
contamination begin to dominate the photometry. Our priors on the different
parameters are set by a variety of considerations. For the RGB tip
magnitude we use a uniform prior over the range F814W=26.5--27.5 (\ie
$m_{\rm tip}=U(26.5,27.5)$); the rapid decline in counts brighter than F814W
$\approx$ 27 suggests this is the approximate location of the tip, and
this is also the expected tip magnitude for old stellar populations
roughly at the distance of the Virgo Cluster. To assign the prior on the
slope of the luminosity function fainter than the tip, we rely on studies of the RGB
slope in nearby galaxies \citep{Mendez02,Makarov06} which generally find a
logarithmic slope of 0.3, and hence adopt a Gaussian prior on the RGB
slope of $a=G(0.3,0.1)$. For the background contamination fraction, we
use our direct measure of contamination on the image ($f_c=0.18$), and
adopt a Gaussian prior of $f_c=G(0.18,0.1)$ to allow for some variation
due to spatial inhomogeneity in the background counts. The slope of the
counts brighter than the RGB tip is a bit more problematic; these
objects could arise from a variety of sources with ill-constrained
luminosity functions, such as AGB stars, chance blends, and uncorrected
contamination. For this portion of the luminosity function we allow a
wide uniform prior on the slope of $c=U(0,2)$. Finally, for the
discontinuity across the RGB tip (the ``break strength'') we assign
another wide uniform prior of $b=U(0,1)$.

\begin{figure*}[]
\centerline{\includegraphics[width=7.0truein]{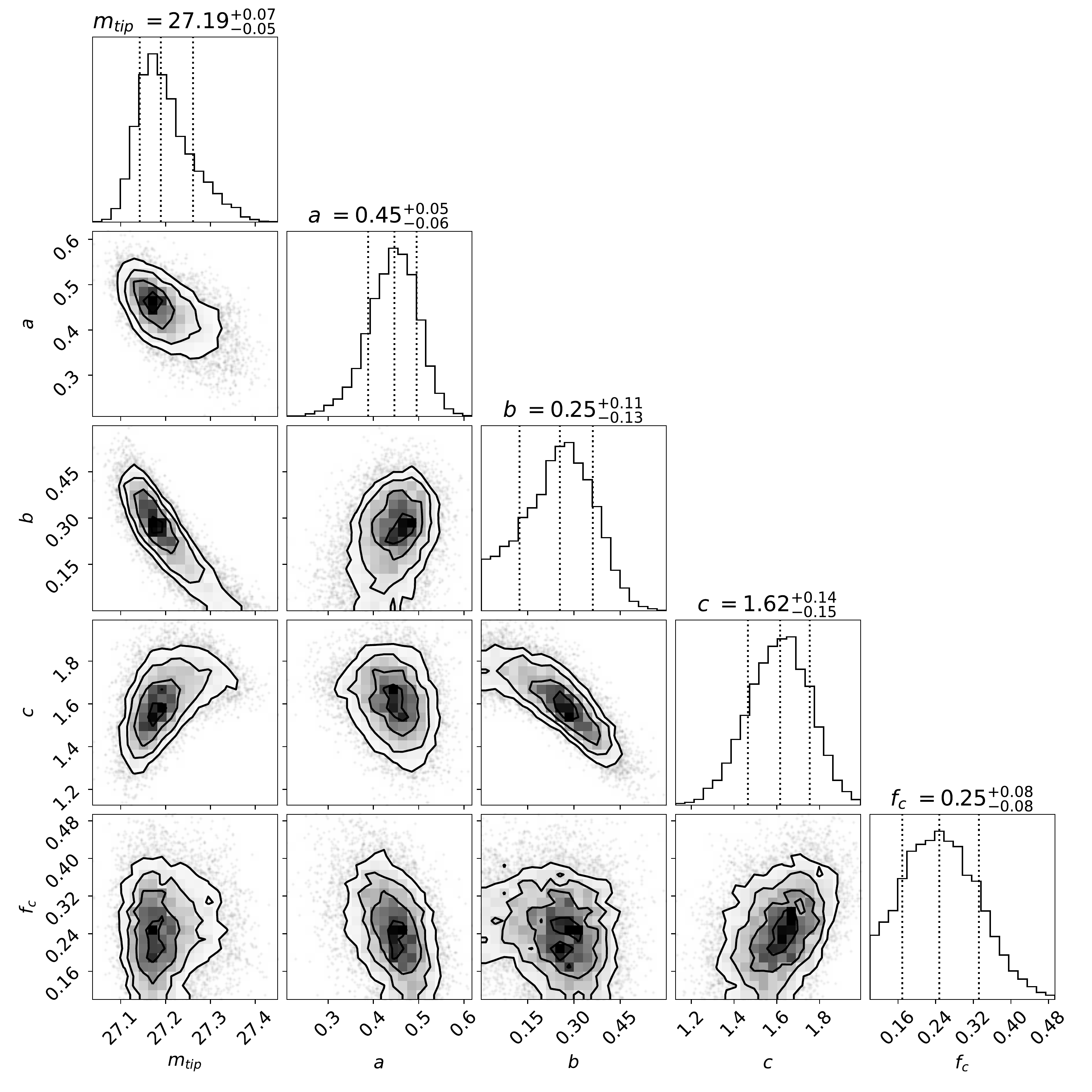}}
\caption{Posterior probability density profiles for the model luminosity
function parameters: TRGB magnitude ($m_{\rm tip}$), luminosity
function slope below the tip ($a$), TRGB break strength ($b$), bright end luminosity
function slope ($c$), and background contamination fraction ($f_c$).
Dotted lines show the median and 16th/84th percentiles of the
marginalized distributions.}
\label{TRGBpars}
\end{figure*}

To estimate the RGB tip magnitude we use the Markov Chain Monte Carlo code 
{\tt emcee v3} \citep{emcee} to sample parameter space for the 
modeled luminosity function and marginalize over the remaining parameters.
We initialize the walkers using the prior distributions for the parameters, burn in
the sampler for 1500 samples ($\approx$ 20 autocorrelation times), and then
run an additional 5000 samples for the final analysis, the results of which
are summarized in Figure~\ref{TRGBpars}. During this analysis, we adopt a
foreground extinction correction of $A_{\rm F814W}=0.04$ \citep{Schlafly11},
and the RGB tip magnitudes reported below include that extinction
correction.

Marginalized over the other parameters, our final estimate for the TRGB
is $m_{\rm tip,F814W}=27.19_{-0.05}^{+0.07}$, where the error bars refer to
the 84th/16th percentiles of the distribution. The posterior
distribution on $m_{\rm tip}$ shows a clear peak, but with an asymmetric
tail towards fainter values. An examination of the covariances between
the parameters in Figure~\ref{TRGBpars} shows that this tail is due to
models with weaker break strengths (lower values for $b$) and steeper
slopes for the bright end of the luminosity function (larger values for
$c$).

\begin{figure*}[]
\centerline{\includegraphics[width=7.0truein]{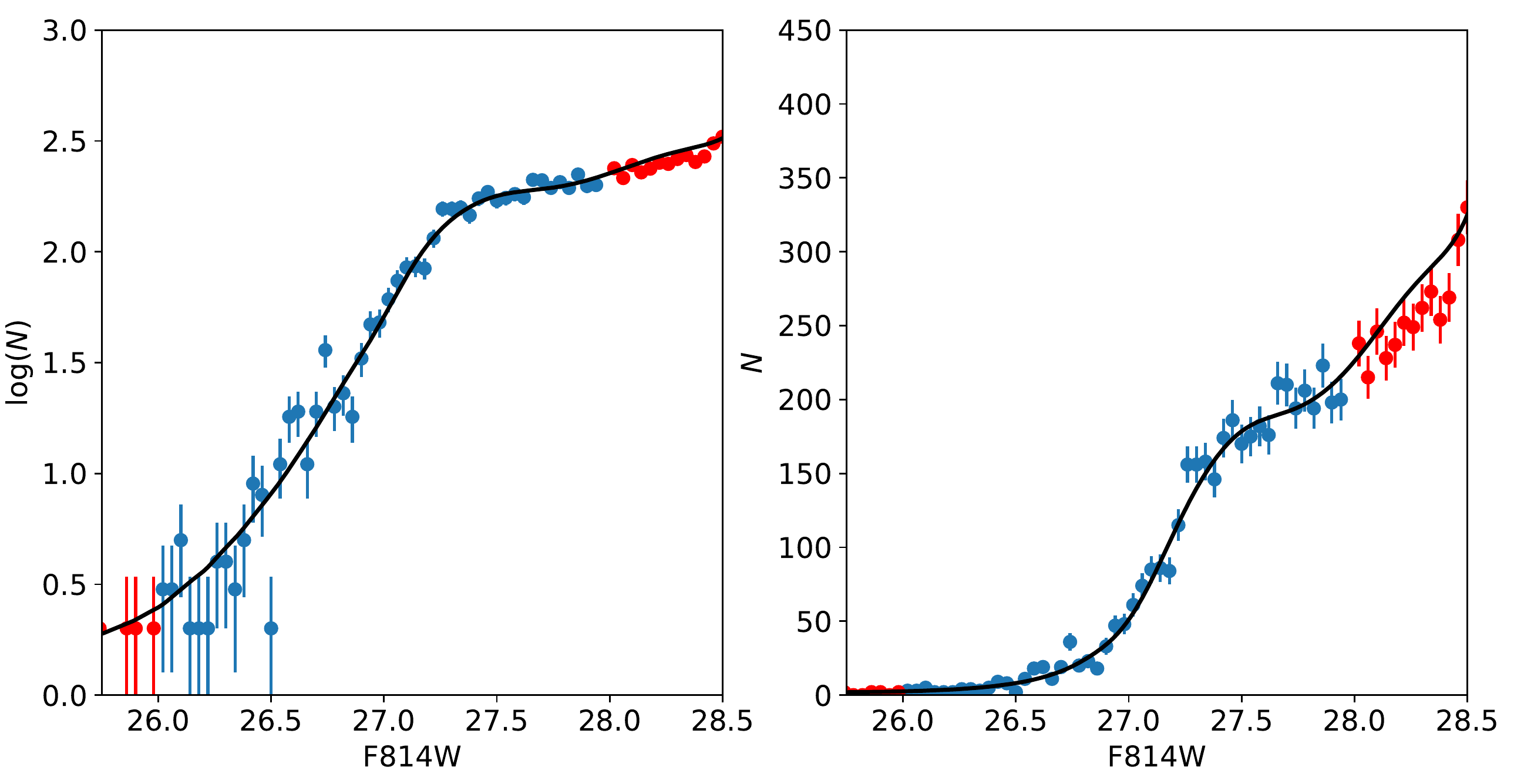}}
\caption{The best fit luminosity function model (black line) overlaid on
the binned observed luminosity function. Red points show magnitude
ranges excluded from the MCMC analysis.}
\label{LFfit}
\end{figure*}

We also examine how our results depend on the radial range chosen for
the analysis. In principle, this choice could lead to systematic
differences in the derived TRGB, for a variety of reasons, including
stellar population gradients within VCC~615 or an increased
contamination fraction in the galaxy's low density outskirts. We break
the detected point sources within VCC~615 into two radially separate
samples: an inner sample covering $r=$ 5\arcsec--30\arcsec\ and an outer
sample at $r=$ 30\arcsec--65\arcsec, with each having roughly similar
numbers of stars. We then run the analysis on each sample separately to
look for systematic differences in the derived parameters of the
underlying luminosity function. These results are shown in
Table~\ref{LFfits}.

Comparing the inner and outer samples, we see a slight change in the
inferred RGB tip magnitude of $\Delta m_{\rm tip}=0.1$, albeit only at
the $1\sigma$ level. If real, this difference may hint at a slight
metallicity gradient in the galaxy -- while the absolute magnitude of
the RGB tip in F814W is relatively constant at metallicities lower than
$[M/H]=-0.7$ \citep{Serenelli17,Beaton18}, models suggest slight
systematic trends between $M_{\rm tip}$ and $[M/H]$ may still be present
at the 0.1 mag level. The general trend is for the tip magnitude to
become fainter as metallicity increases
\citep[\eg][]{Rizzi07,Serenelli17,Jang17}, a more metal-poor outer disk
would thus show a slightly brighter tip magnitude, like that suggested
by our analysis. But at even lower metallicity ($[M/H]<-1$), models
suggest this trend may reverse, with the tip magnitude becoming fainter
as metallicity continues to decrease. Given its low luminosity, VCC~615
likely is quite metal-poor; placing it on the luminosity-metallicity
relationship \citep{Skillman89, Grebel03, Kirby13}, suggests it has a
metallicity near $[M/H]\approx-1$, but without a firm estimate of its
metallicity, and given the uncertainties in both the models and the
data, we leave any intrinsic metallicity gradient as a possible
systematic for further study.

However, another plausible explanation for the slight difference in the
inferred tip magnitude between the inner and outer samples is the effect
of contamination from intracluster RGB stars within the Virgo Cluster
itself. Our inferred tip magnitude ($m_{\rm tip}=27.19$, derived from
the full sample) places VCC~615 slightly behind the cluster core. A
population of intracluster stars concentrated in the Virgo Core would
act as foreground contaminants with slightly brighter tip magnitude.
Since the fractional contamination would be worse in the outskirts of
VCC~615, where the galaxy's stellar density is lower, such contamination
would manifest as the outskirts having a systematically brighter tip
magnitude and higher inferred contamination fraction compared to the
inner sample of stars, precisely as seen in Table~\ref{LFfits}.

Finally, we also look at the systematic effects of extending the
analysis to fainter magnitude ranges, down to F814W=28.3 or 28.5. At
these magnitudes, the stellar incompleteness fraction is high (only 20\%
(12\%) of stars at F814W$=28.3\ (28.5)$ are recovered in our artificial
star tests), and the counts start to become dominated by background
contaminants (see Figure~\ref{lumfuncs}). However, this gives us an
opportunity to better constrain how background contamination
may be diluting and influencing the derived tip magnitude, and also explore any systematic
behavior between the estimated contamination fraction and other inferred
parameters of the fitted luminosity function. The results of this
analysis are also shown in Table~\ref{LFfits}, where it can be seen that
the derived tip magnitude has little dependency on the magnitude range
of the analysis. Extending the analysis to fainter magnitudes does lead
to slight increases in the derived contamination fraction ($f_c$), but
this comes with a corresponding decrease in the inferred RGB slope
(lower values for $a$). In essence, at faint magnitudes, the MCMC
analysis is trading off steeper RGB slopes for a higher contamination
fraction, but with little impact on the parameters at the brighter end
of the luminosity function. Given the low stellar completeness and the
uncertainties in the background contamination, we do not consider these
degenerate shifts in the faint end parameters of the luminosity function
to be meaningful.

In summary, these various tests show that the derived luminosity
function parameters (and most importantly the RGB tip magnitude, $m_{\rm
tip}$) are not highly sensitive to the radial or magnitude range of the
photometric sample being analysed. While there is some hint that factors
such as metallicity or contamination from intracluster RGB stars may
possibly affect our estimate of $m_{tip}$, the evidence is weak, only at
the $1\sigma$ level. Without additional information to constrain the
metallicity of the stellar population or the presence of intracluster
RGB stars, we simply leave this as an open systematic and use the tip
magnitude of $m_{\rm tip}=27.19^{+0.07}_{-0.05}$ (estimated over the
range $r=5\arcsec$--$65\arcsec$ and $F814W=26.0$--$28.0$) as our best
estimate of the RGB tip magnitude in VCC~615. Adopting the
\citet{Freedman20} calibration of the absolute magnitude of the RGB tip
($M_{\rm tip,F814W}=-4.054\pm0.022\ \rm{(stat)} \pm 0.039\ \rm{(sys)}$)
for old metal-poor stellar populations, we derive a final distance
estimate of $d=17.7^{+0.6}_{-0.4}$ Mpc, where the uncertainties include
the statistical uncertainty in $M_{\rm tip,F814W}$. This distance places
VCC~615 slightly behind the Virgo core, at a distance of $d_{\rm
Virgo}=16.5\pm0.1$ Mpc \citepalias{Mei07, Blakeslee09}.

The second, complementary comparison of VCC~615's location with respect
to the Virgo core comes from comparing our VCC~615 TRGB estimate to that
for intracluster stars in the Virgo core by \citet{Lee16}, based on
observations by \citet{Williams07}. \citet{Lee16} find a RGB tip
magnitude of $m_{\rm tip,I} = 26.93 \pm 0.03$. Including the small
$-0.007$ magnitude correction to convert $I$ to F814W
\citep{Freedman19}, we get a difference in the F814W tip magnitudes of
$\Delta m = m_{\rm tip,VCC615} - m_{\rm tip,Virgo} = +0.33 \pm 0.07$, or
a relative distance ratio of $d_{\rm VCC615}/d_{\rm Virgo} =
1.17\pm0.04$, again putting VCC~615 on the far side of Virgo.

As one final consistency check on the nature of the sources in our
VCC~615, we also calculate the ratio of detected stellar sources to the
total broadband light and compare to the expectation from stellar
population synthesis models. Within a radius of $r=50\arcsec$, we
measure a total of $N_*=2380$ RGB stars between $m_{tip}$ and $m_{\rm
tip}+0.5$, after correcting for spatial incompleteness, photometric
incompleteness, and background contamination. Using the $g$-band
integrated light profile from {\sl NGVS} data ($r_e=26.33\arcsec$,
$n=0.67$, \brackmueg=26.86; \citealt{Lim20}), and adopting our TRGB
distance of 17.7 Mpc, this gives a ratio of $N_*/L_{g,\sun} =
5.7\times10^{-5}$ within $r=50\arcsec$. The PARSEC population synthesis
models of \citet{Bressan12} predict values in the range $N_*/L_{g,\sun}
= 5.5-7\times10^{-5}$, for populations spanning a range of metallicities
$[M/H]=-0.7$ to $-2$ and ages 6 to 10 Gyr, in good agreement with our
measured value.

\begin{deluxetable*}{cccccccc}
\tabletypesize{\scriptsize}
\tablewidth{0pt}
\tablecaption{MCMC Modeling Results}
\tablehead{
\colhead{Region} & \colhead{Radial Range} & \colhead{$N_*$} & \colhead{$m_{\rm tip}$} & \colhead{$a$} & \colhead{$b$} & \colhead{$c$} & \colhead{$f_c$}
}
\startdata
\multicolumn{8}{c}{Magnitude Range: 26.0 -- 28.0}\\
\tableline
\bf All & \bf 5\arcsec--65\arcsec & 4177 & $\bf 27.19^{+0.07}_{-0.05}$ & $\bf 0.45^{+0.05}_{-0.06}$ & $\bf 0.26^{+0.11}_{-0.13}$ & $\bf 1.62^{+0.14}_{-0.15}$   & $\bf 0.25^{+0.09}_{-0.08}$\\
Inner & 5\arcsec--30\arcsec & 2179 & $27.27^{+0.09}_{-0.08}$ & $0.42^{+0.07}_{-0.08}$ & $0.18^{+0.13}_{-0.12}$ & $1.64^{+0.12}_{-0.13}$   & $0.20^{+0.08}_{-0.06}$ \\
Outer & 30\arcsec--65\arcsec & 1998 & $27.17^{+0.09}_{-0.06}$ & $0.43^{+0.07}_{-0.07}$ & $0.32^{+0.16}_{-0.18}$ & $1.60^{+0.20}_{-0.24}$   & $0.32^{+0.09}_{-0.09}$ \\
\tableline
\multicolumn{8}{c}{Magnitude Range: 26.0 -- 28.3}\\
\tableline
All & 5\arcsec--65\arcsec & 5941 & $27.20^{+0.07}_{-0.04}$ & $0.38^{+0.04}_{-0.05}$ & $0.25^{+0.11}_{-0.13}$ & $1.64^{+0.13}_{-0.15}$   & $0.29^{+0.08}_{-0.08}$ \\
Inner & 5\arcsec--30\arcsec & 3002 & $27.28^{+0.08}_{-0.07}$ & $0.37^{+0.05}_{-0.06}$ & $0.19^{+0.13}_{-0.13}$ & $1.63^{+0.12}_{-0.13}$   & $0.21^{+0.08}_{-0.07}$ \\
Outer & 30\arcsec--65\arcsec  & 2939 & $27.16^{+0.07}_{-0.05}$ & $0.43^{+0.05}_{-0.06}$ & $0.33^{+0.15}_{-0.17}$ & $1.59^{+0.21}_{-0.24}$   & $0.36^{+0.08}_{-0.09}$ \\
\tableline
\multicolumn{8}{c}{Magnitude Range: 26.0 -- 28.5}\\
\tableline
All & 5\arcsec--65\arcsec  & 7261 & $27.24^{+0.08}_{-0.05}$ & $0.32^{+0.04}_{-0.05}$ & $0.21^{+0.12}_{-0.13}$ & $1.67^{+0.11}_{-0.13}$   & $0.33^{+0.05}_{-0.05}$ \\
Inner & 5\arcsec--30\arcsec & 3588 & $27.30^{+0.08}_{-0.07}$ & $0.32^{+0.06}_{-0.06}$ & $0.17^{+0.13}_{-0.11}$ & $1.66^{+0.11}_{-0.11}$   & $0.26^{+0.06}_{-0.06}$ \\
Outer & 30\arcsec--65\arcsec & 3673 & $27.18^{+0.09}_{-0.06}$ & $0.38^{+0.05}_{-0.06}$ & $0.31^{+0.16}_{-0.18}$ & $1.62^{+0.19}_{-0.24}$   & $0.37^{+0.06}_{-0.06}$ 
\enddata
\tablecomments{The marginalized parameters of the VCC~615 RGB luminosity function derived for photometric samples with
differing radial ranges and magnitude ranges. $N_*$ shows the number of objects
in each sample. 
$m_{\rm tip}$ is the F814W magnitude of the RGB
tip, $a$ and $c$ are the logarithmic slope of the LF fainter and brighter than the tip magnitude, respectively, $c$
is the logarithmic discontinuity in counts at the tip, and $f_c$ is the contamination fraction measured at F814W=28.0. 
Error bars in the derived parameters show the 16th/84th percentile ranges.. Our adopted model solution is shown in boldface. See text for details.}
\label{LFfits}
\end{deluxetable*}

\section{Discussion}

With a best fit TRGB distance to VCC~615 of $d=17.7^{+0.6}_{-0.4}$ Mpc,
our analysis places the galaxy $1.2^{+0.6}_{-0.5}$ Mpc on the far side
of the Virgo Cluster. Its three dimensional position relative to M87
gives it a Virgocentric radius of $\approx 1.3$ Mpc, close to the
cluster's virial radius ($R_{200}=1.5$ Mpc; \citealt{McLaughlin99}).
This placement of VCC~615 on the far side of Virgo is also supported by
its fainter RGB tip magnitude compared to intracluster red giants
measured by \citet{Lee16}, and any weak systematic effects we could
plausibly identify in our analysis would push the galaxy even further
beyond the Virgo core. The galaxy's measured line-of-sight velocity
($v_{\rm VCC615}=2094\pm3$ \kms; \citealt{Toloba18}) is redshifted by
$\approx$ 1000 \kms compared to the Virgo mean velocity ($v_{\rm
Virgo}=1088\pm105$ \kms; \citetalias{Mei07}). Projecting the relative
line of sight velocity onto the Virgo radius vector gives a radial
velocity with respect to the cluster of $875 \pm 140$ \kms, where the
uncertainty reflects both the VCC~615 distance uncertainty and the
uncertainty in the cluster mean distance and velocity. Of course, the
unknown velocity in the sky plane will affect Virgocentric radial
velocity as well, but since the largest component of the radius vector
is along our line of sight, the true Virgocentric radial velocity is
unlikely to be much different from that calculated here.

\begin{figure}[]
\centerline{\includegraphics[width=3.2truein]{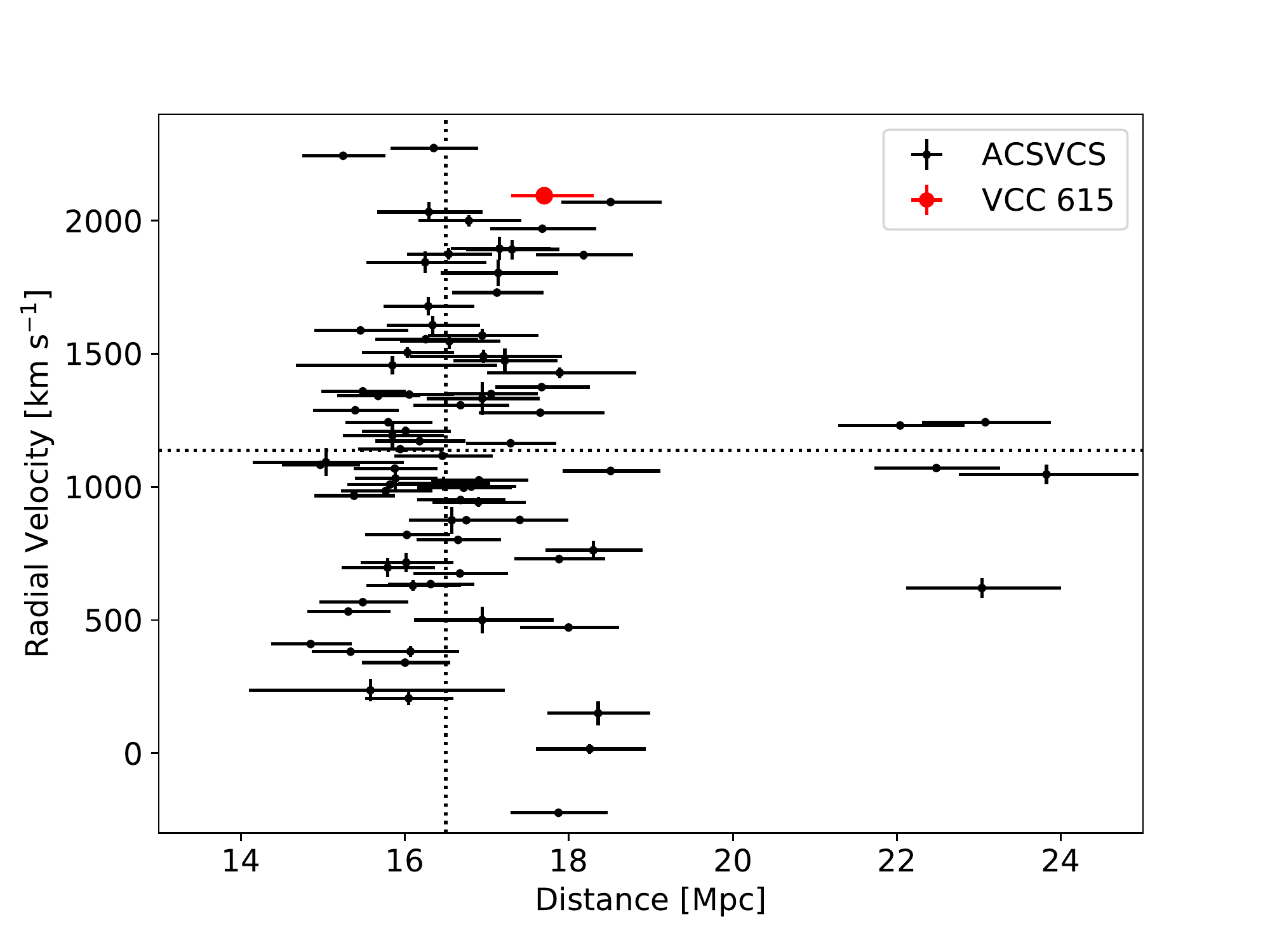}}
\caption{Velocity-distance plot for galaxies in Virgo. Black points are
galaxies from the ACS Virgo Cluster Survey, with distances derived from
surface brightness fluctuations \citepalias{Mei07, Blakeslee09}. The red
point shows our TRGB distance to VCC~615, along with its measured radial
velocity from \citet{Toloba18}. Grey dotted lines show the mean Virgo
distance (16.5 Mpc) and velocity (1138 \kms) from \citetalias{Mei07,
Blakeslee09}. }
\label{VirgoPV}
\end{figure}

While VCC~615's radial velocity is somewhat high for Virgo cluster
member galaxies, it is not anomalously so. At the galaxy's
position within Virgo, the escape velocity from the main cluster (using
the mass model of \citealt{McLaughlin99}) is $\approx$ 2400 \kms. 
Unless the unknown transverse velocity is significantly higher than the
galaxy's Virgocentric radial velocity of 875 \kms, VCC~615 is thus likely a bound member
of the Virgo cluster. Figure~\ref{VirgoPV} places VCC~615 on the
distance-velocity diagram for other Virgo galaxies determined using
surface brightness fluctuation distances from \citetalias{Blakeslee09}.
Galaxies within Virgo have a large velocity spread; the
main cluster (Virgo A) has a velocity dispersion of
$\sigma_{v_r}=593\pm68$ \citep{Mei07}, but the full velocity range is
much higher: $-220\ {\rm km\ s^{-1}} \leq v_r \leq +2300\ {\rm km\
s^{-1}}$ \citep[see also][]{Binggeli87}. In Figure~\ref{VirgoPV},
VCC~615 follows the general distribution of other Virgo galaxies, and we
also see no evidence for VCC~615 being part of any of Virgo's
kinematically distinct substructure. Projected 1.9\degr\ WSW of M87, the
galaxy does not lie within any of the boundaries of the Virgo M,
W$^\prime$ or W clouds (as defined by \citealt{Binggeli87}), nor is its
distance or radial velocity consistent with those subgroups.

Thus the combined position-velocity data for VCC~615 are consistent with
the galaxy being a kinematically normal member of the main Virgo
cluster. Furthermore, the galaxy is currently on an outbound trajectory,
leaving the inner parts of the cluster. Under the simplest assumption of a
free-flight line-of-sight trajectory (\ie no transverse velocity), the galaxy would have experienced
Virgo pericenter 1.3 Gyr ago, having passed 0.5 Mpc from the cluster
center. Of course, since a free-flight calculation disregards the effects
of the cluster potential, it overestimates the time since periapse passage.
Nonetheless, it is clear that VCC~615 is not on an inbound trajectory, and 
if it is an object that recently fell into the cluster for the first time, it has had
many dynamical times to respond to the effects of the cluster potential.
Given the galaxy's half-light radius (2.2 kpc) and the velocity dispersion 
of its globular clusters ($\sigma_v=32^{+17}_{-10}$ \kms; \citealt{Toloba18}),
the galaxy has a rough dynamical time of $(G\rho)^{-1/2} \approx 70$ Myr,
much larger than the crossing time of the cluster.

Without knowing VCC~615's transverse velocity, it is difficult to make
any more definitive statement about its orbital trajectory through
Virgo, but with some assumptions about the cluster potential and
kinematics, we can make probablistic statements about the dynamical
history of the galaxy. To do this, we adopt the NFW mass model of
\citet{McLaughlin99} for Virgo Subcluster A (the main mass component of
Virgo), which, scaled to a distance of 16.5 Mpc has the parameters
$R_{200}=1.55$ Mpc, $R_s=0.62$ Mpc, and $M_{200}=4.2\times10^{14}
M_{\odot}$. Given VCC~615's three-dimensional position relative to M87
(which we take to define the cluster center), we then assume orbital
isotropy, calculate the 1-d velocity dispersion of the NFW profile at
that radius \citep{Lokas01}, and draw the two missing components of
VCC~615's velocity vector from a Gaussian probability distribution given
that dispersion. We then use the software package {\tt gala}
\citep{gala} to integrate the orbit backwards in the model
potential to derive the distance and time of last pericenter passage
(\rperi, \tperi). We do this for 10,000 randomly sampled trajectories
initialized in this fashion to study a plausible distribution of \rperi\
and \tperi\ for VCC~615.

\begin{figure*}[]
\centerline{\includegraphics[width=7.0truein]{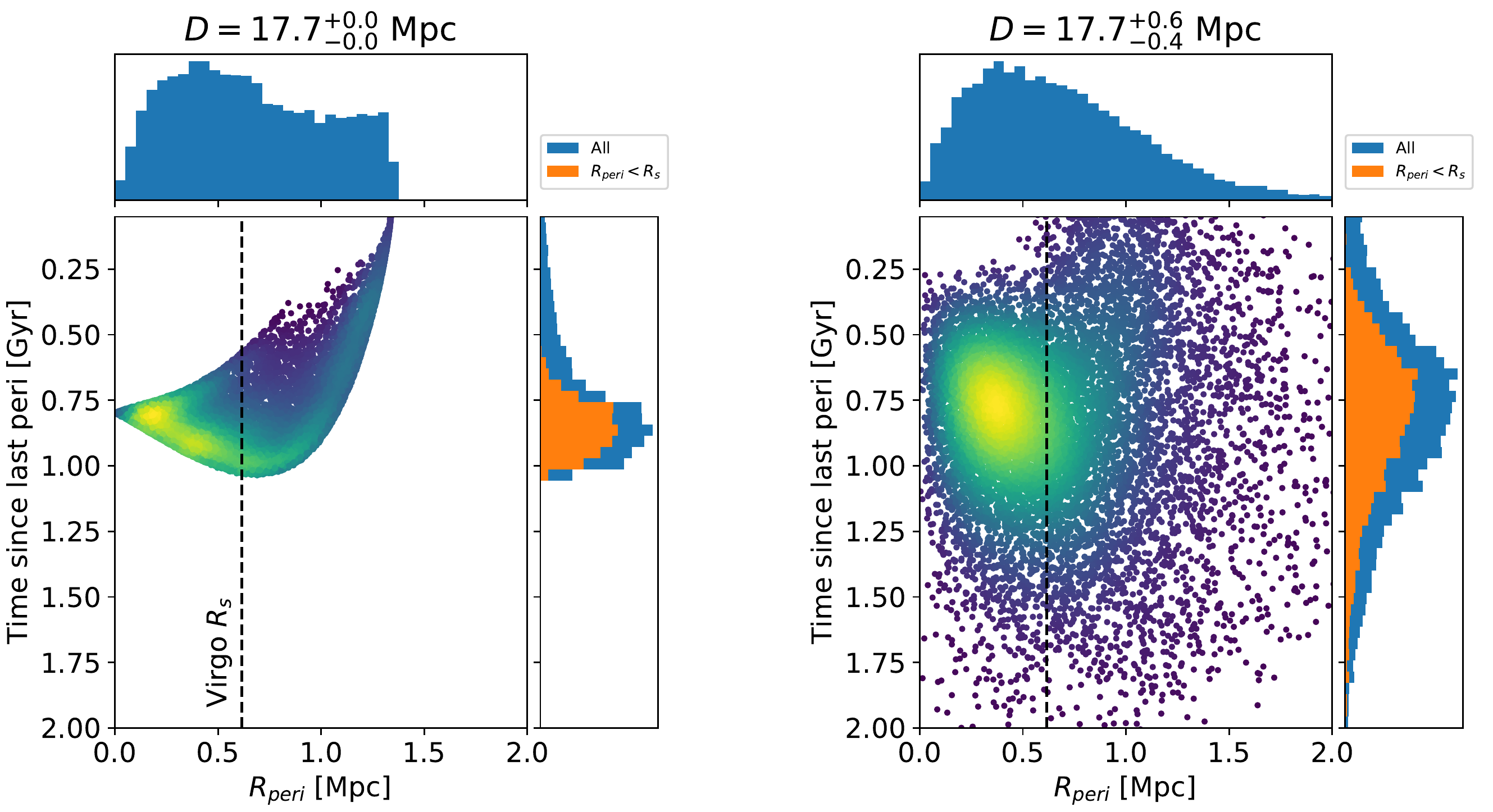}}
\caption{Inferred probablistic orbit distribution for VCC~615, showing
the distribution of time and distance of the most recent Virgocentric periapse for
orbits which sample a range of tangential motions drawn from the Virgo
kinematic model described in the text. The left panel shows orbits which
start with VCC~615 precisely at our best-match distance and assuming no
uncertainty in the adopted Virgo distance and mean radial velocity. The
right panel repeats the analysis, but starts VCC~615 at a range of
distances randomly sampled from our uncertainty estimates, and also
incorporated the uncertainties in the Virgo distance and velocity. In
the main plot, the colormap shows the density of points, while in the
histograms, the orange histogram shows the distribution of periapse time
for orbits that reach inside the Virgo Cluster scale radius. See text
for full details.
}
\label{orbits}
\end{figure*}

The results of this calculation are shown in Figure~\ref{orbits}. The
left panel shows the results for an ensemble of orbits which all start
precisely at our best match distance of $D=17.7$ Mpc and presume no
uncertainties on the distance and mean velocity of the cluster. While
this makes for an overly idealized calculation, it is instructive for
understanding the orbital behavior of the models. Purely radial orbits
are those with \rperi=0, passing through the cluster center $\approx$
800 Myr ago. At the other extreme are orbits which maximize the
tangential velocity; these are now at closest approach (\tperi=0) and
have high velocity across the plane of the sky that is starting to carry
them away from pericenter. Between those extremes, at a given
$R_{peri}$, orbits with the most recent pericenter passage have high
space motion and carry the galaxy quickly through the cluster on orbits
that extend well outside $R_{200}$, while orbits with older pericenter
times are more tightly bound, with the galaxy moving more slowly, with
little transverse space motion. The dashed vertical line in
Figure~\ref{orbits} shows the scale radius for \citet{McLaughlin99} NFW
model describing the main body of Virgo(Virgo A; $R_s=0.62$ Mpc), used
here as a metric defining the cluster core. Half (49.9\%) of the orbits
in the ensemble travel inside $R_s$, with pericenter passages typically
happening $\approx$ 750 -- 900 Myr ago. While many orbits carry the
galaxy inside $R_s$, few orbits pass truly deep into the cluster core;
only 10.1\% of the orbits reach a pericenter closer than 200 kpc of the
cluster center.

The right panel of Figure~\ref{orbits} more accurately captures
uncertainties in the orbital parameters, by factoring in the uncertainty
in VCC~615's distance, as well as in the Virgo distance and velocity,
when initializing the orbits. To capture the velocity uncertainty we
start the orbits with a line-of-sight velocity for VCC~615 drawn from a
Gaussian with mean $\Delta v = v_{\rm VCC615} - v_{\rm Virgo}$ and
dispersion equal to the uncertainty in the Virgo mean velocity ($\pm
105$ \kms, \citet{Mei07}). The uncertainty in position is modeled by
sampling the posterior distribution function for our TRGB distance shown
in Figure~\ref{TRGBpars}. These variations smear out the derived orbital
parameters, but the fraction of orbits where the galaxy passes inside
$R_s$ (49.3\%) or inside 200 kpc (9.4\%) are both essentially unchanged.
The time since pericenter passage for these orbits peaks around 750 Myr,
but shows a large distribution, with the 25th and 75th percentile range
covering the range 640$-$1030 Myr. Of course, even this statistical
analysis of orbits contains a number of dynamical assumptions embedded
in it: that the missing components of the tangential velocity can be
modeled using the NFW distribution for the Virgo potential; that VCC~615
has not experienced strong interactions with other galaxies in Virgo;
and that the influence of other Virgo mass components (such as the Virgo
B clump to the south) is negligible. Nonetheless, the broad implications
seem robust: {\it VCC~615 is on an outbound trajectory from Virgo,
having experienced pericenter passage at least 500 Myr or more in the
past.}

VCC~615's outbound orbit shows that rather than being fragile objects
relegated to the cluster outskirts, at least some cluster UDGs can and
do survive passages through the dense cluster environment. While the
uncertainties in the orbit modeling make it ill-advised to project the
orbit too far back in time, the galaxy has likely experienced some
amount of dynamical processing by the cluster potential; nearly 50\% of
the orbit projections have it at or inside the cluster core within the
past 1.5 Gyr. However, given the VCC~615's dynamical mass and
mass-to-light ratio ($M_{\rm dyn}=2.7\times10^9 M_\odot, (M/L)_V=56
M_\odot/L_\odot$; \citealt{Toloba18}, both measured inside the globular
cluster half light radius, and also scaled for our updated distance),
the galaxy is likely quite resilient against the cluster tidal field. At
its current Virgocentric distance ($R_{\rm Virgo}=1.3$ Mpc), VCC~615's
tidal radius ($r_t=R_{\rm Virgo}(M_{\rm VCC615} / M_{\rm Virgo})^{1/3}$)
is $\sim$ 15 kpc, and even within the cluster core the tidal radius
remains significantly higher than the galaxy's effective radius (2.2
kpc). To suffer serious tidal disruption, VCC~615 would need to pass
quite close to the cluster center ($\lesssim$ 200 kpc), and our orbit
modeling suggests that passages this close are unlikely.

Of course, the \citet{Toloba18} mass estimate, based on its globular
cluster kinematics, presumes that the globulars are fair kinematic
tracers of the VCC~615's dynamical mass. While the galaxy is currently
in the outskirts of the cluster where tides should not strongly
influence the globular cluster kinematics, a recent core passage could
have dynamically heated the clusters and yield an inflated measure of
the galaxy's mass. For example, if the mass-to-light ratio of the galaxy
was more similar to that of normal Virgo dE's ($M/L_V \sim$ 3--5,
\citealt{Toloba14}), its binding mass would be reduced by an order of
magnitude. In this case, inside of $R_s$ the galaxy's tidal radius would
be reduced to 5 kpc or less, comparable to its effective radius. Thus,
in principle, a close core passage could have dynamically heated the
galaxy, leading to both its diffuse nature and the high velocity
dispersion of its globular cluster system. However, VCC~615 shows no
morphological sign of tidal heating --- it does not appear tidally
distorted, nor do we see any evidence for stellar tidal streams in the
galaxy's outskirts. Thus, while the orbital modeling indicates that an
recent core passage for VCC~615 is quite possible, we think it unlikely
that such an event has strongly affected the galaxy's structure or
kinematics.

In the context of cluster UDG evolution models, the properties of
VCC~615 seem in many ways consistent with the ``born-UDG'' galaxy
populations in the simulations of \citet{Sales20}. In these simulations
born-UDGs form as normal LSB galaxies in the field and are later
accreted into the cluster, in contrast to ``tidal-UDGs'' which fell into
the cluster early and have been dynamically heated and stripped by the
cluster tidal field. Born-UDGs tend to be found in the cluster
outskirts, like VCC~615, and have don't show evidence for suppressed
velocity dispersions indicative of significant tidal mass loss.
VCC~615's high velocity dispersion \citep{Toloba18} and lack of tidal
deformation are consistent with the expectation for these born-UDGs.
Nonetheless, VCC~615's outbound trajectory and the orbit models shown in
Figure~\ref{orbits} indicate that the galaxy has been deeper in the
cluster potential in the past, where it has likely experienced some
amount of tidal processing. Its long term survival in the cluster is
then likely due to its massive dark matter halo ($M_{\rm halo} \approx
10^{11}$~-- $10^{12}\ M_{\odot}$, \citealt{Toloba18}) protecting it from
complete tidal disruption. Its very high velocity dispersion also makes
it an outlier in galaxy dynamical scaling relationships
\citep{Toloba18}, suggesting it is not merely a normal LSB galaxy that
fell into Virgo; instead, some additional process must have driven its
evolution into its UDG form. Whether these processes are
cluster-specific (such as the effects of ram-pressure stripping
\citealt{SafScan17}), or occured in the field before infall
\citep[\eg][]{diCintio17, Chan18} remain unclear. Follow-up observations
of the stellar populations and metallicity of VCC~615 will help resolve
these uncertainties.

\section{Summary}

We have used deep {\sl Hubble Space Telescope} imaging to resolve
individual stars within the Virgo ultradiffuse galaxy VCC~615. We use the data
to estimate a distance to the galaxy using the tip of the red giant branch (TRGB) 
distance indicator. We then couple this distance with the galaxy's observed
radial velocity to infer its dynamical history within the Virgo cluster and
compare to formation models of ultradiffuse cluster galaxies. Our main findings
are summarized here.

\begin{enumerate}

\item Our modeling of the RGB luminosity function yields an observed
TRGB magnitude (corrected for foreground extinction) of $m_{\rm
tip,F814W}=27.19^{+0.07}_{-0.05}$. Using the TRGB calibration of
\citet{Freedman20}, this places VCC~615 at a distance of
$d=17.7^{+0.6}_{-0.4}$ Mpc.

\item Our TRGB distance puts VCC~615 on the far side of the Virgo
Cluster ($d_{\rm Virgo}=16.5$ Mpc; \citetalias{Mei07, Blakeslee09}).
This inference is further supported by the $\Delta m_{\rm tip} \approx
+0.3$ magnitude difference between the TRGB magnitude reported here for
VCC~615 and that derived for intracluster RGB stars in the Virgo core
\citep{Lee16}.

\item Given the galaxy's three-dimensional position inside Virgo,
VCC~615 lies at a Virgocentric radius of 1.3 Mpc, in the outskirts close
to the cluster's virial radius. Coupled with its observed radial
velocity (2094 \kms; \citealt{Toloba18}), we find the galaxy is moving
away from the Virgo core at relatively high velocity ($\Delta v = 1000$
\kms). This, however, is within the normal range of Virgocentric
velocities, and thus the combined position--velocity data shows VCC~615
to be a member of the main Virgo cluster.

\item Orbit modeling using VCC~615's position--velocity data and the
Virgo mass model of \citet{McLaughlin99} indicates a $\sim$ 50\% chance
that the galaxy passed through the Virgo core ($r<620$ kpc) in the past
Gyr, but only $\sim$ 10\% of orbits take the galaxy deep ($r<200$ kpc)
into the cluster center.

\item The spatial distribution of RGB stars within VCC~615 shows no
evidence of tidal distortion, and we see no evidence of tidal streams
around the galaxy. If VCC~615 did in fact have a recent core passage,
cluster tides appear not to have strongly affected the galaxy's
morphology.

\item Comparing to formation scenarios for cluster UDGs, VCC~615's
outbound trajectory from Virgo shows that it is not an infalling field
UDG. While it may have experienced a recent core passage, it is unlikely
to have been {\it very} deep in the cluster potential; combined with its
morphologically undisturbed nature, this argues against a recent
transformation from a low mass dwarf into a cluster UDG. Instead, it is
more likely that the galaxy is a long-lived member of the Virgo cluster,
where its dynamical mass is sufficient to prevent rapid stripping and
tidal disruption. If the cluster environment has led to VCC~615's status
as a UDG, that process likely happened long ago.

\item From the spatial density of point sources in the field surrounding
VCC~615 we put an upper limit on the local surface brightness of the
Virgo intracluster light of $\mu_B > 30.1$ mag arcsec$^{-2}$, assuming
an old (10 Gyr), metal-poor ([M/H]=$-2$) stellar population. After
correcting this limit for likely contamination due to unresolved
background sources, our best estimate for the surface brightness of the
intracluster light is $\mu_B = 30.5$ mag arcsec$^{-2}$. These estimates
are consistent with deep imaging of the Virgo Cluster \citep{Mihos17},
which detected no intracluster light around VCC~615 down to a limiting
surface brightness of $\mu_{B,{\rm lim}} = 29.5$ mag arcsec$^{-2}$.

\item Finally, we also report the discovery of smaller, diffuse galaxy
$\approx$ 90\arcsec\ away from VCC~615. The luminosity function of
resolved stars within this galaxy is consistent with it being a member
of the Virgo Cluster. The galaxy has an effective radius
$r_e=5.1$\arcsec\ (400 pc), integrated magnitude $m_g=21.8\ (M_g=-9.2)$
and mean effective surface brightness of \brackmueg=27.4 \magsec, where
the numbers in parentheses give the physical properties at a distance of
16.5 Mpc.

\end{enumerate}

\acknowledgments

We would like to thank Andy Dolphin for his guidance in performing
point-source photometry with DOLPHOT, as well as Jay Anderson with help
and suggestions regarding slightly trailed HST images. We also thank the
anonymous referee for a helpful and lightning-fast referee report. This research is
based on observations made with the NASA/ESA Hubble Space Telescope for
program \#GO-15258 and obtained at the Space Telescope Science Institute
(STScI). STScI is operated by the Association of Universities for
Research in Astronomy, Inc., under NASA contract NAS5-26555. Support for
this program was provided by NASA through grants to J.C.M. and P.R.D.
from STScI. LVS is grateful for support from the NSF-CAREER-1945310 and
NASA ATP-80NSSC20K0566 grants. 

\facility{HST (ACS), CFHT}
\software{
astropy  \citep{astropy}, 
numpy \citep{numpy},
matplotlib \citep{matplotlib},
scipy \citep{scipy},
emcee \citep{emcee},
gala \citep{gala}
}
  
\bibliographystyle{aasjournal}

\end{document}